\documentclass[twocolumn,prd,preprintnumbers,amsmath,amssymb]{revtex4}

\usepackage{graphicx}
\usepackage{dcolumn}
\usepackage{bm}
\usepackage{epsfig}
\usepackage{amssymb}
\usepackage{amsmath}

 \def\tskip{\setlength{\tskip}{5pt}}
\def\colwidth{\setlength{\colwidth}{3.5in}}

\def\prd{Phys. Rev. D}

\def\prl{Phys. Rev. Lett.~}

\def\apj{Astrophys. J.~}

\def\mnras{Mon. Not. Roy. Astron. Soc.~}

\def\be{\begin{equation}}
\def\ee{\end{equation}}
\def\ba{\begin{eqnarray}}
\def\ea{\end{eqnarray}}

\newcommand{\lsim}{\mathrel{\hbox{\rlap{\lower.55ex\hbox{$\sim$}} \kern-.3em \raise.4ex \hbox{$<$}}}}
\newcommand{\gsim}{\mathrel{\hbox{\rlap{\lower.55ex\hbox{$\sim$}} \kern-.3em \raise.4ex \hbox{$>$}}}}
\newcommand{\beq}{\begin{equation}}
\newcommand{\eeq}{\end{equation}}
\newcommand{\beqa}{\begin{eqnarray}}
\newcommand{\eeqa}{\end{eqnarray}}

\begin{document}

\title{Gravitational-Wave by Binary OJ287 in 3.5 PN Approximation}

\author{Y. Zhang\footnote{yzh@ustc.edu.cn}, S. G. Wu and W. Zhao\footnote{wzhao7@ustc.edu.cn}}
\affiliation{ Key Laboratory for Researches in Galaxies and
Cosmology, Department of Astronomy, University of Science and
Technology of China, Hefei, Anhui, 230026, China}

\date{\today}

\begin{abstract}

We compute the gravitational radiation of the supermassive binary
black hole OJ287. By fitting the data of its recent seven
outbursts, we obtain the orbital motion up to 3.5PN
(Post-Newtonian) order, the energy and angular momentum fluxes and
the  waveform up to the 2PN. It is found that the 1PN term of the
energy flux, has an opposite sign to the Newtonian flux, and the
2PN term has an opposite sign to the 1PN one. The same pattern is
also found for the angular momentum fluxes. The total flux  is
reduced by $30.8\%$ on average from the Newtonian flux for OJ287,
resulting in a smaller orbital decay rate than that of 2.5PN
calculation. Consisting with this, it is checked that, in the
sequence of non-dissipative PN forces, each term of $i^{th}$ order
(dissipative or non-dissipative) has a sign opposite to the
$(i-1)^{th}$ order. The origin of this characteristic is traced to
the appearance of the non-diagonal metric component $g_{0i}$ in PN
approximation. This feature will have profound impact on
estimation of gravitational waves from binary systems.

\end{abstract}


\maketitle



~\\
\textbf{\emph{Introduction}} Blasar OJ287 at a  redshift  $ z =
0.306$ has long been observed continuously, and has been
identified as a binary supermassive black hole system
\cite{Sillanp88,Sundelius1997,Valtonen2006,Valtonen0609,Valtonen2007}
with masses $m_1=1.84\times10^{10}M_{\odot}$ and
$m_2=1.46\times10^{8}M_{\odot}$. Having a very high orbital
velocity $v/c\simeq (0.06\sim  0.26)$, it serves as a test
platform for general relativity up to 3PN \cite{Valtonen2011},
mainly via the precession effect on the orbit. Nevertheless, OJ287
is also supermassive, its gravitational radiation is very strong
with amplitude $h\sim 10^{-15}$ as we shall see, and is a proper
target of the low-frequency gravitational wave detectors. In this
paper we compute its orbital motion up to 3.5PN, i.e, including
the second order of dissipative term. Then we obtain for the first
time its gravitational waveform up to 2PN, and demonstrate the
energy and angular momentum fluxes up to 2PN correction to the
Newtonian fluxes. The result is comprehensive and can be used for
waveform templates for detectors.


~\\ \textbf{\emph{Orbital Motion to 3.5PN}} Since $m_1\gg m_2$,
the configuration of the binary is modelled as such that the
secondary is orbiting around the primary, the accretion disk of
the primary is perpendicular to the orbital plane
\cite{Valtonen2006,Valtonen0609,Valtonen2007}. We do not include
the possible spin effects for simplicity \cite{ValtonenSpin}. The
3.5PN equation of motion in the center-of-mass frame is given by
\cite{IyerWill93,IyerWill95,Jaranowski97,PatiWill02,
KonigFayeSchafer03,Blanchet02,NissankeBlanchet05},\ba
\label{equation} \frac{d\mathbf{v}}{dt}&=&-\frac{Gm}{r^{2}}\times
\nonumber
\\ &\times&[(1+A_{\rm{1PN}}+A_{\rm{2PN}}+A_{\rm{2.5PN}}
+A_{\rm{3PN}}+A_{\rm{3.5PN}}){\mathbf e}_{r} \nonumber \\
&+&(B_{\rm{1PN}}+B_{\rm{2PN}}+B_{\rm{2.5PN}}+B_{\rm{3PN}}+B_{\rm{3.5PN}})\mathbf{v}]
, \ea where $m=m_1+m_2$, $\bf v$ is the relative velocity, $\bf
e_r$ is the radial unit vector, $A_{\rm{iPN}}$ and $B_{\rm{iPN}}$
are the PN corrections of the  $i$ order. In the polar coordinate
$(r,\Phi)$ on the orbital plane, the solution is obtained
straightforwardly \cite{WuZhang}, given an initial condition:
$r_0$, $\dot r_0$, $\Phi_0$, $\dot \Phi_0$. By fitting with the
observational data of the last seven outbursts of OJ287
\cite{Sundelius1997,Valtonen2006,Valtonen0609}, and taking into
account of the time delay  effect  \cite{WuZhang}, using the
method of least squares for the difference between the predicted
and observed times of seven outbursts, we arrived at the optimal
orbital parameters \cite{WuZhang},
\begin{center}
\begin{tabular}{ll}
 \hline\\[-4.5mm]
 $m_1$ & $~~~~1.84\times10^{10}M_{\odot}$ \\
 $m_2$ &$~~~~1.46\times10^{8}M_{\odot}$ \\
 ${\rm period\,} P$  & $~~~~ 11.9248$ years  \\
 ${\rm ellipticity\, }e$   & $~~~~ 0.6693$ \\
 ${\rm semi-major\,  axis\, }a$   & $~~~~ 0.0535 $  pc  \\
 $\omega$ & $~~~~238.35^{\circ}$ at 1971.48  \\
 ${\rm precession\, rate\,}\Psi$   & $~~~~32.6359^{\circ}$/period \\
\hline\\[-4.5mm]
\end{tabular}
\end{center}

\begin{figure}[t]
\begin{center}
\includegraphics[width=8cm]{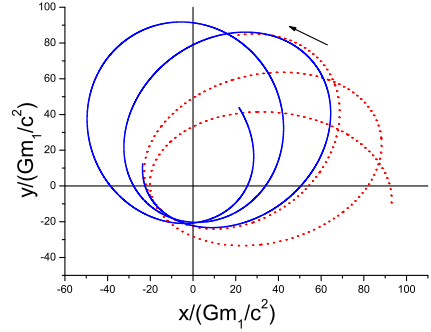}
\caption{  \label{orbit} After $\sim$200 periods, 3.5PN orbit
(solid) surpasses that of 2.5PN (dash), both starting with the
same initial position. }
\end{center}
\end{figure}

\begin{figure}[t]
\begin{center}
\includegraphics[height=6cm, width=9cm]{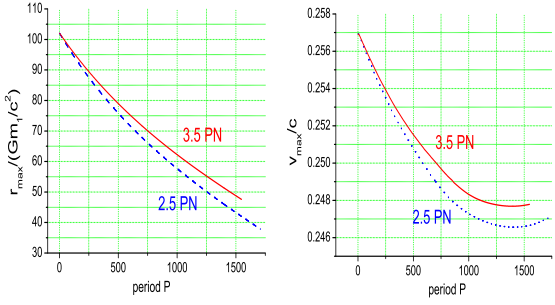}
\caption{  \label{rvmax} The maximum mutual separation $r_{\max}$
(left) and speed $v_{\max}$ (right) in 3.5PN are decreasing more
slowly than those in 2.5PN, respectively.
 }
\end{center}
\end{figure}

The overall behavior of orbital motion is that the secondary is
orbiting around with a very large $\Psi$ due to the PN forces, the
orbit is continuously shrinking and becomes more circular in the
course of time, due to the reaction of GW radiation. We have
checked that the 3.5PN orbital motion fits the data better than
the 2.5PN one by the least squares
\cite{Sundelius1997,Valtonen2007}. The detail has been given in
Ref.\cite{WuZhang}. As Fig.\ref{orbit} shows, the 3.5PN orbit has
a slightly greater precession and gradually surpasses the 2.5PN
orbit. More interestingly, as shown in Fig.\ref{rvmax}, the mutual
separation and velocity in the 3.5PN calculation are shrinking
more slowly than those in the 2.5PN one, indicating that the
binary described by 2.5PN must radiate away more gravitational
energy than the 3.5PN description. Now we investigate  this
feature in detail.

~\\ \textbf{\emph{The Signs of PN Forces}} The first dissipative
terms are at 2.5PN ones \cite{IyerWill95,Blanchet02}, \ba
\label{A25} A_{\rm{2.5PN}}&=& \frac{1}{c^{5}} \nu
\frac{Gm}{r}\dot{r}
    \left( -\frac{24 v^{2}}{5} - \frac{136}{15}\frac{Gm}{r}
    \right), \\
\label{B25} B_{\rm{2.5PN}}&=&\frac{1}{c^{5}}\nu \frac{Gm}{r}
       \left(  \frac{8 v^{2}}{5}+   \frac{24}{5}\frac{Gm}{r}  \right),
\ea with $\nu=m_1m_2/m^2$. They cause damping of the orbit and
give rise to the Newtonian quadrupole radiation. The next order
dissipative terms are  3.5PN, \ba \label{A35} A_{\rm{3.5PN}}&=&
\frac{1}{c^{7}}\nu \frac{Gm}{r} \dot{r} \bigg[ (\frac{366
v^{4}}{35}+ 12 \nu v^{4}  \nonumber \\ &-&114v^{2}
\dot{r}^{2}-12\nu v^{2}\dot{r}^{2}+ 112 \dot{r}^{4}) \nonumber \\
\nonumber
           &+&\frac{Gm }{r }(\frac{692  v^{2}}{35}-
\frac{724 v^{2}\nu }{15}+\frac{294 \dot{r}^{2}}{5}+ \frac{376\nu
\dot{r}^{2}}{5})  \\ &+& \frac{G^2 m^2 }{r^2} (\frac{3956
}{35}+\frac{184 \nu }{5})   \bigg], \ea

\ba  \label{B35} B_{\rm{3.5PN}}&=&\frac{\nu}{c^{7}} \frac{Gm}{r}
\bigg[  (-\frac{626 v^{4}}{35} \nonumber \\ &-&\frac{12 \nu
v^{4}}{5}+\frac{678 v^{2}\dot{r}^{2}}{5} +\frac{12\nu
v^{2}\dot{r}^{2}}{5}  -120 \dot{r}^{4}) \nonumber \\
          &+&\frac{Gm}{r} (\frac{164  v^{2}}{21}
+\frac{148\nu  v^{2}}{5}-\frac{82  \dot{r}^{2}}{3} -\frac{848\nu
\dot{r}^{2}}{15}) \nonumber \\
 &+&\frac{G^2 m^2 }{r^2}(-\frac{1060 }{21}-\frac{104\nu }{5})\bigg].
\ea Without these four dissipative terms, as has been checked, the
3PN orbit will precess only and the radius will not shrink, so
that the energy and angular momentum are conserved.

\begin{figure}
\begin{center}
\includegraphics[height=7cm,width=9.5cm]{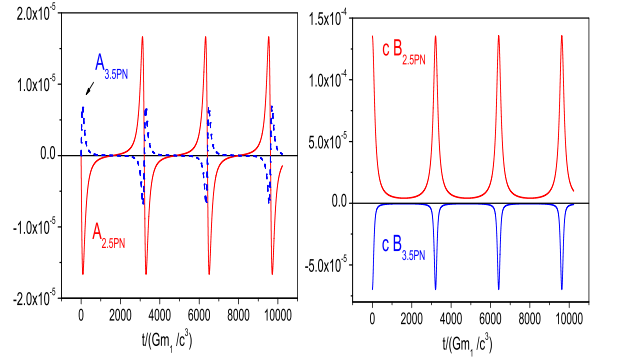}
\caption{  \label{fig4}
 Left: $A_{\rm{2.5PN}}$ and $A_{\rm{3.5PN}}$ have opposite signs at any instance.
Right: $B_{\rm{2.5PN}}>0$ and $B_{\rm{3.5PN}}<0$. }
\end{center}
\end{figure}

Fig.\ref{fig4} demonstrates that, $A_{\rm{2.5PN}}$ and
$A_{\rm{3.5PN}}$ always have opposite signs at any instance of
time, and the amplitude of  $A_{\rm{3.5PN}}$ is $\sim 0.3$ that of
$A_{\rm{2.5PN}}$. So is with $B_{\rm{2.5PN}}$ and
$B_{\rm{3.5PN}}$. Therefore, the 3.5PN force is just in opposite
direction to the 2.5PN one, the 3.5PN description yields less
orbital damping and less gravitational radiation than the 2.5PN
one. Similarly, we find that in the sequence of the
non-dissipative PN terms, $A_{\rm{1PN}}$, $A_{\rm{2PN}}$,
$A_{\rm{3PN}}$, each term has a sign just opposite to its
precedent term at any instance of time. The same alternating
behavior also occurs for the sequence of $B_{\rm{1PN}}$,
$B_{\rm{2PN}}$, $B_{\rm{3PN}}$.

\begin{figure}
\begin{center}
\includegraphics[height=9cm, width=9cm]{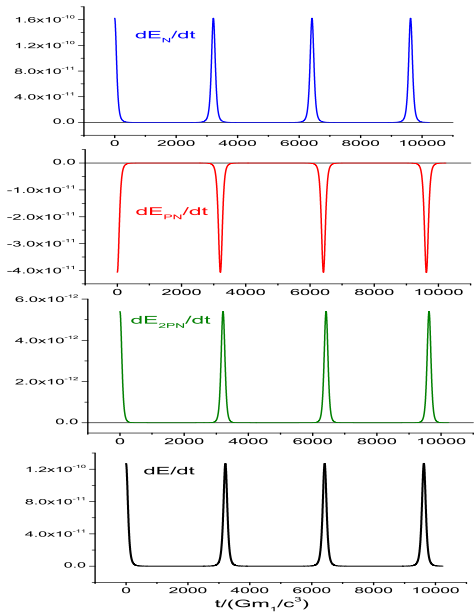}
\caption{   \label{fig6} The energy fluxes (in unit  $c^5/G$)
$\dot E_{\rm N}>0$, $\dot E_{\rm{1PN}}<0$, $\dot E_{\rm{2PN}}>0$.
}
\end{center}
\end{figure}

\begin{figure}
\begin{center}
\includegraphics[width=9cm]{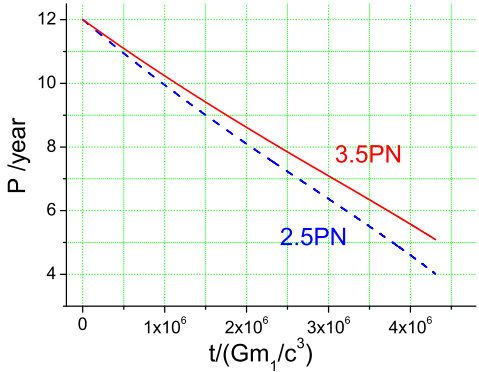}
\caption{   \label{period} The period $P$  in 3.5PN decreases at a
slower rate than in 2.5PN. }
\end{center}
\end{figure}

~\\ \textbf{\emph{The Energy and Angular Momentum Fluxes}} Since
the dissipative 3.5PN forces are opposite to the 2.5PN, the work
done $(A_{\rm{3.5PN}}  {\bf e_r} + B_{\rm{3.5PN}}  {\bf v})\cdot d
{\bf r} $ is consequently opposite to $(A_{\rm{2.5PN}} {\bf e_r} +
B_{\rm{2.5PN}}  {\bf v})\cdot d {\bf r} $. Thus one expects that,
in the total energy flux
\cite{BlanchetShafer89,WillWiseman96,Gopakumar2} \be  \label{dEdt}
\frac{dE}{dt}= \dot E_{\rm N} + \dot E_{\rm{1PN}} +  \dot
E_{\rm{2PN}}, \ee the dominant Newtonian flux \be \dot E_{\rm N} =
\frac{8}{15} \frac{G^3m^4\nu^2}{c^5r^4}  (12v^2-11\dot r^2) \ee
induced by 2.5PN force has an opposite sign to \ba \dot
E_{\rm{1PN}} &=& \frac{8}{15}
\frac{G^3m^4\nu^2}{c^5r^4}\frac{1}{28c^2} \times \nonumber \\&
&[-2 (1487-1392\nu) v^2 \dot r^2
-160(17-\nu)\frac{m}{r}v^2  \nonumber \\
& & +(785-852\nu)v^4
+3(687-620\nu)\dot r^4  \nonumber \\
& & +8(367-15\nu) \frac{m}{r} \dot r^2 +16(1-4\nu)(\frac{m}{r})^2
] \ea induced by 3.5PN force. In turn, for the same reason, $\dot
E_{\rm{2PN}} $ (see Refs. \cite{WillWiseman96,Gopakumar2}) induced
by 4.5PN force \cite{Gopakumar1} has an opposite sign to  $\dot
E_{\rm{1PN}} $ \cite{WillWiseman96}. The energy fluxes of OJ287
are explicitly demonstrated in Fig.\ref{fig6} with $\dot E_{\rm N}
>0$, $\dot E_{\rm{1PN}} <0$, $\dot E_{\rm{2PN}}>0 $, which is in a complete
agreement with the relative signs of 3.5PN and 2.5PN forces.
Similarly, the angular momentum fluxes
\cite{BlanchetShafer89,Junker,Gopakumar2} ${dJ}/{dt}= \dot J_{\rm
N} + \dot J_{\rm{1PN}} +  \dot J_{\rm{2PN}}$ of OJ287 is computed
with a result that $\dot J_N
>0$, $\dot J_{\rm{1PN}}<0$, and $\dot J_{\rm{2PN}}>0$. During each
orbiting period the emission occurs mostly around the periastron
where $r$ is minimum and $v$ attains the maximum. Integrated over
a period, the negative $\dot E_{\rm{1PN}} $ reduces $dE/dt$ by
$\sim 30\%$, and $\dot J_{\rm{1PN}} $ reduces $dJ/dt$ by $\sim
17\%$. This huge amount of reduction of radiated energy and
angular momentum is due to the high orbital speed of OJ287. Fig.
\ref{period} shows that the period $P$ decreases more slowly in
3.5PN  than in 2.5PN, a result that is explained by $\dot
E_{\rm{1PN}} <0$ and $\dot J_{\rm{1PN}}<0$. This pattern is
consistent with the formula (4.29) of $P$ and the formula (4.31)
of $\dot P$ in Ref.\cite{Gopakumar2}, but different from (4.26) in
Ref.\cite{BlanchetShafer89}. The lesson learned is that,
 2.5PN calculations  of
the Newtonian fluxes $\dot E_N$,  $\dot J_N$, and the heuristic
rate  $\dot P$ \cite{Perters-Mathews} are over-estimations. A
similar pattern has also been noticed in gravitational recoil due
to linear momentum loss, where the Newtonian, 1PN, and 2PN
contributions  change sign alternatingly
\cite{Fitchett,Wiseman,Blanchet}.



~\\ \textbf{\emph{The Waveform}} Up to 2PN, the
traceless-transverse (TT) gravitational waveform is given by
\cite{WillWiseman96} \be \label{hii} h^{ij}_{TT}=\frac{2G m\nu}{R}
   [ Q^{ij} + P^{1/2}Q^{ij} + P Q^{ij} +  P^{3/2} Q^{ij} + P^2
   Q^{ij}]_{TT} \nonumber
 \ee
where $R$ is the distance from the binary to the observer,
$Q^{ij}$ is the dominant, quadrupole momentum part, and the
remaining terms with superscripts denote the the effective PN
order (their explicit expressions are given by (6.11) in
Ref.\cite{WillWiseman96}). Using the resulting orbital motion, we
compute $h^{ii}_{TT}$ up to 2PN order. Since OJ287 is at cosmic
distance with $z\sim 0.306$, the effect of cosmic expansion should
be taken into account. The distance $R$ is taken to be the
luminosity distance: \be D_L(z)= \frac{c}{H_0}
        (1+z)\int_0^z \frac{dz'}{ \sqrt{\Omega_{m}(1+z')^3+ \Omega_{\Lambda} }}
\ee in a spatially flat $\Lambda$CDM  Universe with
$\Omega_{\Lambda}= 0.685$, $H_0=67.3$(km/s)/Mpc and
$\Omega_{m}=1-\Omega_{\Lambda}$ \cite{planck}, where $H_0$ is the
Hubble
constant. For OJ287, $R=D_L(0.306)\simeq 1646.89$ Mpc. 
The prefactor amplitude  of $h_{ij}$ is $\frac{2G m\nu}{R} \simeq
8.42\times 10^{-15}$. Let the observer's direction be given by
${\bf n} = \sin\theta {\bf e_x} +\cos\theta {\bf e_z}$, where $\bf
e_z$ is the normal vector and $\bf e_x$ is along the semimajor
axis on the orbital plane. Take  an orthonormal basis (${\bf
e_\theta}$, ${\bf e_\phi}$, ${\bf n}$). The two independent
components of GW measured by the observer are \cite{WagonerWill76}
\ba \label{hij} h^{\theta\theta}_{TT} &=& -  h^{\phi\phi}_{TT}
           = \frac{1}{2}(\cos^2\theta h^{xx}-h^{yy}),
\\ h^{\theta\phi}_{TT} &=& \cos\theta h^{xy}. \ea Fig.\ref{hij} shows
the wave form $h^{\theta\theta }_{TT}$ and $h^{\theta\phi }_{TT}$
using the 3.5PN orbital result. The peak amplitude reaches
$|h_{ij}| \sim 1\times 10^{-15}$ with a frequency $f \sim
2.66\times 10^{-9}$ Hz, higher than the amplitude of relic
gravitational waves in this frequency range \cite{Zhang}. Both can
be the targets of pulsar timing arrays, such as PPTA  \cite{PPTA},
EPTA \cite{EPTA}, NANOGrav \cite{NANOgrav}, FAST \cite{fast} and
SKA \cite{SKA}. But its frequency may be too low for LIGO and
VIRGO \cite{LIGOS5}.

\begin{figure}
\begin{center}
\includegraphics[height=8cm, width=9cm]{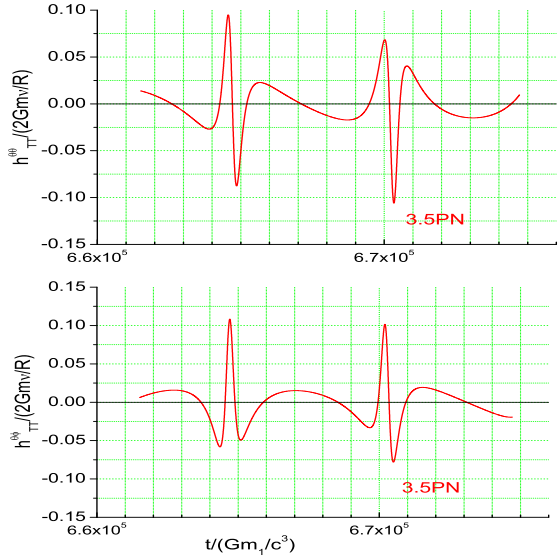}
\caption{   \label{hij} The waveform $h^{\theta\theta}_{TT}$
(upper) and  $h^{\theta\phi}_{TT}$ (lower) in 3.5PN. The observer
is facing the orbital plane with the angle $\theta=0$. }
\end{center}
\end{figure}


~\\ \textbf{\emph{The Origin of the Opposite Sign}} Now let us see
the origin of the opposite sign occurring at each next order.
Inspect that,  aside the common factor, $A_{\rm{2.5PN}}$ in
Eq.(\ref{A25}) contains the negative factor $( -\frac{24 v^{2}}{5}
- \frac{136}{15}\frac{Gm}{r}  )$, while the dominant terms  of
$A_{\rm{3.5PN}}$  in  Eq.(\ref{A35}), such as $v^4$, $\dot r ^4$,
$v^2 \frac{G m }{r}$, $ \dot{r}^2 \frac{G m }{r}$, $ ( \frac{G m
}{r} ) ^2$, are positive. Thus $A_{\rm{3.5PN}}$ has an opposite
sign to $A_{\rm{2.5PN}}$. Similarly, inspection of Eq.(\ref{B25})
tells that $B_{\rm{2.5PN}}>0$, while the dominant terms of
$B_{\rm{3.5PN}}$ in Eq.(\ref{B35}), such as $v^4$, $\dot r ^4$, $
\frac{G^2 m^2 }{r^2}$, in square bracket, all have large negative
coefficients, so $B_{\rm{3.5PN}}<0$.

This feature can be inferred generally from the PN approximation,
say, in the method of the PN iterative metric parameterized by
retarded potentials \cite{NissankeBlanchet05}. The acceleration is
determined by $a^i =F^i-\frac{d}{dt} (P^i-v^i)$ (see Eq.(2.13) in
Ref.\cite{NissankeBlanchet05}), where
\[
F^i \sim \partial_i V +\frac{1}{c^2}(-V\partial_i V
+\frac{3}{2}\partial_i V \, v^2  -4\partial_i  V_j \,
v^j)+O(1/c^4),
\]
\[
(P^i-v^i) \sim \frac{1}{c^2} (\frac{1}{2}v^2 v^i +3V v^i -4V^i
)+O(1/c^4),
\]
with $V\sim Gm/r\sim v^2$ and $V_j \sim v^j V$ to leading order.
For each order, the dominant contribution from $\partial_i V $ is
followed by the next order contribution
$\frac{1}{c^2}(-V\partial_i V +\frac{3}{2}\partial_i V \, v^2
-4\partial_i  V_j \, v^j) \sim -\frac{7}{2}\frac{1}{c^2}v^
2\partial _i V$, and $-\frac{d}{dt}(P^i-v^i)
 \sim -\frac{d}{dt}  \frac{1}{c^2}(\frac{1}{2}v^2 v^i -V v^i )
 \sim   \frac{1}{c^2} \frac{d}{dt}  \frac{1}{2} v^2 v^i
 \sim  \frac{1}{c^2} \frac{3}{2}v^2 \partial_i V $.
The sum of these two terms is $ \sim - \frac{1}{c^2}2v^2
\partial_i V $ having an opposite sign to $\partial_i V$. Thus,
for each PN order of $a^i$, the next order has an opposite sign.
The cause is traced to $\frac{1}{c^2}(-4\partial_i  V_j \, v^j)$
in $F^i $, which originates from the metric component
$g_{0i}=-4V_i/c^3+O(1/c^5)$, i.e., the shift
\cite{NissankeBlanchet05}.


~\\ \textbf{\emph{Conclusion}} We have studied gravitational
radiation of the binary black hole OJ287, apply the Post-Newtonian
approximation up to 3.5PN order in the orbital motion. By explicit
computations, we have found that, at any instance of time, the
energy flux $\dot E_{\rm{1PN}}$ induced by the 3.5 force is
negative, opposite to the Newtonian flux $\dot E_{N}$ induced by
the 2.5PN force, and, consistent to this, the dissipative 3.5PN
force is always opposite to the 2.5PN one. For OJ287 this
reduction of energy flux is as high as $\sim 30\%$. Therefore, for
binaries with high orbiting speed, this effect is significant and
must be taken into account in evaluation of radiation of GW. We
have also demonstrated that, for non-dissipative PN forces, each
PN order has a sign opposite to the precedent PN order. The origin
this characteristic has been traced to the non-diagonal metric
component $g_{0i}$ in PN approximation. With a peak amplitude
$|h_{ij}| \sim 10^{-15}$ and low frequencies $f \sim 2.7\times
10^{-9}$ Hz, the GW from OJ287 is an ideal target for pulsar
timing array detectors.

~

{\bf Acknowledgements:} YZ is supported by the NSFC No.11073018,
11275187, SRFDP, and CAS. WZ is supported by NSFC No.11173021,
11075141 and project of Knowledge Innovation Program of CAS.


\end{document}